\documentclass[twocolumn,pra,superscriptaddress,amsmath,amssymb,aps,reprint]{revtex4-2}

\usepackage{graphicx}
\usepackage{amsmath,amsfonts,bbm}
\usepackage{braket}
\usepackage[normalem]{ulem}
\usepackage{IEEEtrantools}
\usepackage{mathrsfs}
\newcommand{\ketbra}[2]{\mathinner{|{#1}\rangle\,\langle{#2}|}}

\begin{document}

\title{Security of Binary-Modulated Optical Key Distribution Against Quantum-Enhanced Coherent Eavesdropping}

\author{Karol Łukanowski}
\email{k.lukanowski@cent.uw.edu.pl (corresponding author)}
\affiliation{Centre for Quantum Optical Technologies, Centre of New Technologies, University of Warsaw, Banacha 2c, 02-097 Warsaw, Poland}

\author{Michał Wójcik}
\affiliation{Faculty of Physics, University of Warsaw, Pasteura 5, 02-093 Warsaw, Poland}

\author{Stefano Olivares}
\email{stefano.olivares@fisica.unimi.it}
\affiliation{Dipartimento di Fisica “Aldo Pontremoli”, Università degli Studi di Milano, via Celoria 16, I-20133 Milan, Italy}
\affiliation{Istituto Nazionale di Fisica Nucleare, Sezione di Milano, I-20133 Milan, Italy}

\author{Konrad Banaszek}
\email{k.banaszek@uw.edu.pl}
\affiliation{Centre for Quantum Optical Technologies, Centre of New Technologies, University of Warsaw, Banacha 2c, 02-097 Warsaw, Poland}
\affiliation{Faculty of Physics, University of Warsaw, Pasteura 5, 02-093 Warsaw, Poland}

\author{Marcin Jarzyna}
\email{m.jarzyna@cent.uw.edu.pl}
\affiliation{Centre for Quantum Optical Technologies, Centre of New Technologies, University of Warsaw, Banacha 2c, 02-097 Warsaw, Poland}

\begin{abstract}
Optical key distribution (OKD) protects the physical layer of communication links by taking advantage of the inherent noise present in the photodetection process. It allows for efficient generation of a shared random key between two distant users that is secure against passive eavesdropping and can be subsequently used for cryptographic purposes. Moreover, it can be straightforwardly implemented over standard intensity modulation and direct detection links, making it an attractive alternative to quantum key distribution. Here we present a comprehensive security analysis against more powerful eavesdroppers possessing the ability to perform either coherent detection, or even quantum-optimal measurements on the intercepted transmission.
\end{abstract}

\maketitle

\section{Introduction}
The physical layer of modern optical communication links remains largely unprotected from third party interference, with the bulk of security delegated to computational encryption in the higher layers of the network stack~\cite{StallingsBook}. Due to the rising risk of hacking attacks against public key cryptosystems, either via improved classical methods or with the possible advent of large-scale quantum computing~\cite{Shor1997}, more attention is directed towards procedures that provide additional safeguards to information transmission by taking advantage of the physics of the communication channel~\cite{Banaszek2025}. One such avenue is quantum key distribution (QKD), in which two distant devices (which we shall customarily call Alice and Bob) perform measurements on exchanged quantum states of light to obtain correlated sequences of bits~\cite{pirandola2020advances}. Due to the no-cloning theorem for quantum states, with proper protocol engineering and digital post-processing, these sequences are assured to be unknown to potential third parties. When in possession of such shared secret sequences, Alice and Bob may equip their protocol with additional information-theoretic private key cryptography, or use them for fresh seeding of the security measures already in place, which strengthens their resilience. However, reliance on quantum phenomena makes QKD difficult to realize experimentally, especially in large-scale communication networks, as the technology required includes sophisticated techniques like single-photon or shot-noise-limited coherent detection and high-fidelity quantum state preparation. Moreover, inevitable signal attenuation and excess noise vastly limit the possible generation rates of quantum keys and the range of QKD links~\cite{Pirandola2017}.

The recently proposed optical key distribution (OKD) protocol~\cite{Ikuta_2016,OKD} relaxes the technological requirements of QKD while offering the possibility of high-speed private key generation protected against passive eavesdropping, which currently is the most likely form of attack. The protocol can be readily implemented in standard intensity modulation \& direct detection systems, as its security is derived only from the inherent randomness of the photodetection process. It is applicable both to optical fiber links~\cite{JachuraOFC2024} and free-space communication~\cite{Czerwinski2026}, and can be integrated with simultaneous standard data transmission~\cite{jachura2023modelling,Lukanowski2025}. On the other hand, while QKD can in principle protect against intercept-and-resend attacks, such as when a third party intercepts, measures, and resends the signal to the receiver device, OKD assumes that any potential third party (Eve) is at best a passive eavesdropper that may capture a part of the signal coming from Alice, but cannot modify the part that reaches Bob. This means that active eavesdropping such as source manipulation, detector blinding, or man-in-the-middle attacks, is not addressed in the present OKD security analysis. On the other hand, the passive eavesdropping model covers the most relevant current threats to communication security~\cite{JachuraOFC2024, Czerwinski2026, KucharczykOECC2026}. In the case of fiber communication, it includes the collection of light leakage by fiber bending or tapping into unused ports in a switch device~\cite{Karlsson2022}. In case of free-space optical communication, it includes the possibility of Eve collecting the signal that arrives outside of the area secured by Bob (e.g., for satellite or deep-space optical transmission, this could entail installing a rogue receiver telescope nearby and collecting the diffracted or scattered light).

In the previous theoretical study~\cite{banaszek2021optimization}, OKD has been shown to guarantee positive key distribution rates if the eavesdropper Eve implements direct detection (DD) on the intercepted signals, same as Bob. Importantly, even if Eve captures the majority of the signal, this simply lowers the rate of key generation but does not preclude it, in contrast to wiretap channel coding~\cite{Bloch2011}. In this work we show that in the asymptotic information-theoretic regime OKD allows for positive secret key generation rates even for the most general passive eavesdropper capabilities allowed by quantum theory. This paper is organized as follows. In Section~\ref{sec:dd} we review the OKD protocol and previous results on the security against DD eavesdropping. Subsequently, we equip Eve with shot-noise-limited coherent detection capabilities in Section~\ref{sec:coh}, the ability to perform Helstrom measurements which minimize the probability of state discrimination error in Section~\ref{sec:helstrom}, and finally with optimal collective measurements that saturate the Holevo bound on accessible information in Section~\ref{sec:holevo}. Section~\ref{sec:conclusion} summarizes the obtained results and concludes the paper. The main text analysis assumes that Bob and Eve measure signals of macroscopic strength, which allows us to approximate their outcome distributions as Gaussian. In the Appendix we drop this assumption and treat numerically an exact model, which is shown to quickly converge to the approximate one with increasing signal strength.

\begin{figure}[t]
\centering
\includegraphics[width=\columnwidth]{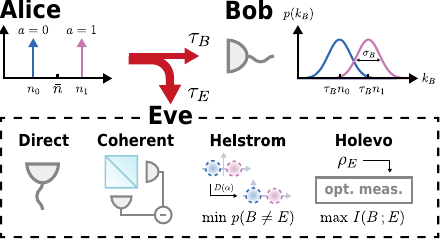}
\caption{Schematic view of the OKD protocol and different eavesdropping scenarios considered in this work. In binary-modulated OKD, Alice prepares one of two macroscopic pulses of similar energy values $n_0$ and $n_1$. She transmits them to Bob via a channel with transmission $\tau_B$. Bob implements direct detection and records normally distributed outputs $k_B$. The eavesdropper Eve passively captures fraction $\tau_E$ of Alice's signal. In the article, four strategies of Eve are considered: direct detection (Sec.~\ref{sec:dd}), coherent detection (Sec.~\ref{sec:coh}), error-minimizing Helstrom measurements (Sec.~\ref{sec:helstrom}, here pictured in reference to the Helstrom-achieving displacement receivers~\cite{Dolinar1973}), and optimal measurements saturating the Holevo bound for accessible information (Sec.~\ref{sec:holevo}). }
\label{fig1}
\end{figure}

\section{Direct detection eavesdropping}
\label{sec:dd}
The goal of the OKD protocol is to establish between Alice and Bob a common sequence of random bits that are unknown to Eve. In each protocol round of binary-modulated OKD, Alice chooses randomly bit value $a \in \{0,1\}$, which constitutes an instance of her random variable $A$, and transmits to Bob a light pulse with one of two possible values of optical energy $n_a \in \{n_0, n_1\}$. The optical channel is characterized by transmission $\tau_B$, with an additional side channel to Eve characterized by transmission $\tau_E$. We assume that the light pulses received by Bob and Eve are of macroscopic strength, i.e., $\tau_B \, n_a \gg 1$ and $\tau_E \, n_a \gg 1$ for both values of $a$. In this regime the discrete Poissonian distribution of photocounts can be well approximated by a Gaussian distribution. Bob's and Eve's detected variables are then modeled by normally distributed continuous variables $k_B | a \sim \mathcal{N}(\tau_{B} \, n_a, \sigma_{B,a}^2)$ and $k_{E} | a \sim \mathcal{N}(\tau_{E} \, n_a, \sigma_{E,a}^2)$, where $\sigma^2_{B/E,\,a}$ denotes the variance of Bob's/Eve's outcome distribution given the value of $a$. Importantly, the energies $n_0$ and $n_1$ are chosen sufficiently close so that neither Bob nor Eve can unambiguously identify the value of each transmitted bit, due to the overlap of photocount distributions as depicted in Fig.~\ref{fig1}. Because for large $n_a$ the photocount outcome variance scales like $n_a$, due to the assumption of small $n_1 - n_0$ difference we further assume that photocount variance depends only on the average $\bar{n} = (n_0 + n_1)/2$ and has the same value for both $a$, $\sigma^2_{B/E,a} = \sigma^2_{B/E}$. While this approximation results in greatly simplified key rate formulas, we further verify its validity in the Appendix, where an exact model with $a$-dependent variances is shown numerically to quickly converge to the present one with equalized variances.

The rate of key generation $\mathsf{K}$ attainable from the raw data held by Alice and Bob after completing the protocol is upper bounded by the Csiszár-Körner theorem valid for individual attacks~\cite{Cover2005},
\begin{eqnarray}
\label{eq:K_ind}
    \mathsf{K} &=& \max \{I(B\,;A) - I(B\,;E), \; 0\} \nonumber\\
    &=& \max \{ H(B|E) - H(B|A), 0\},
    \end{eqnarray}
expressed either as the difference of mutual information between Alice's and Bob's variables $I(B\,;A)$ and between Bob's and Eve's $I(B\,;E)$, or the difference of respective conditional Shannon entropies $H(B|E)$ and $H(B|A)$. Note that this bound is an information-theoretic expression valid in the asymptotic regime of the number of protocol rounds and neglects finite-size corrections~\cite{Portmann2022} as well as other detrimental factors that may limit the key rate achievable in practice, such as imperfect key reconciliation efficiency between Alice and Bob~\cite{ElkoussQIC2010}.

To calculate the key rate it is convenient to first shift and rescale Bob's and Eve's variables,
\begin{equation}
\label{eq:bob_eve_variables}
    y_B = (k_B - \tau_B \bar{n})/\sigma_B,\quad y_E = (k_E - \tau_E \bar{n})/\sigma_E,
\end{equation}
where $\bar{n} = (n_0 + n_1)/2$. We then find the distributions of $y_B$ and $y_E$ conditioned on $A$:
\begin{equation}
\label{eq:yB_yE}
    y_B|a \sim \mathcal{N}\left((-1)^{1-a} \,\delta_B, 1\right), \quad y_E|a \sim \mathcal{N}\left((-1)^{1-a} \,\delta_E, 1\right),
\end{equation}
where 
\begin{equation}
    \delta_B = \frac{\tau_B (n_1 - n_0)}{2 \sigma_B},\quad \delta_E = \frac{\tau_E (n_1 - n_0)}{2 \sigma_E}.
\end{equation}
The conditional entropy $H(B|A)$ can be simply obtained as the differential entropy of a standard normal distribution,
\begin{equation}
\label{eq:HBA}
    H(B|A) = \frac12 \log_2(2 \pi e).
\end{equation}
The calculation of $H(B|E)$ is more involved due to the need to condition Bob's variable on Eve's knowledge. In general,
\begin{align}
\label{eq:HBE_int}
    H(B|E) &= - \int_{{\mathbbm R}^2}\!\!\! \mathrm{d}y_E \, \mathrm{d}y_B \, p(y_B, y_E) \log_2 p(y_B | y_E),
\end{align}
where $p(y_B, y_E)$ is the joint probability distribution of Bob's and Eve's variables and $p(y_B | y_E)$ is the probability distribution of Bob's variable conditioned on Eve's. For $y_B$ and $y_E$ defined according to Eq.~\eqref{eq:bob_eve_variables}, the integral in Eq.~\eqref{eq:HBE_int} can be simplified to
\begin{multline}
\label{eq:HBE}
    H(B|E) = \frac12 \log_2 (2 \pi e) + \delta_B^2 \log_2 e \\- \int_{{\mathbbm R}^2}\!\!\! \mathrm{d}y_E \, \mathrm{d}y_B \, p(y_B, y_E) \log_2\!\frac{\cosh(\delta_E y_E + \delta_B y_B)}{\cosh(\delta_E y_E)}
\end{multline}
with
\begin{multline}
    \label{eq:joint_DD_distribution}
    p(y_B, y_E) = \frac{1}{4 \pi} \left[\exp\left(-\frac{(y_B + \delta_B)^2 + (y_E + \delta_E)^2}{2}\right) \right.  \\ \left. + \exp \left(-\frac{(y_B - \delta_B)^2 + (y_E - \delta_E)^2}{2}\right) \right ]\!.
\end{multline}

Introducing the \emph{eavesdropper's advantage} parameter 
\begin{equation}
\mathscr{E} = \left(\frac{ \tau_E / \sigma_E}{ \tau_B / \sigma_B}\right)^2,
\end{equation}
one may relate
\begin{equation}
\label{eq:deltas_relation}
    \delta_E / \delta_B = \sqrt{\mathscr{E}}.
\end{equation}
Therefore, one may express the key rate in Eq.~\eqref{eq:K_ind} as a function only of the eavesdropper's advantage $\mathscr{E}$, which is defined by the channel parameters, and either the $\delta_B$ or the $\delta_E$ parameter, which depend on $n_1$ and $n_0$ and thus are determined by Alice's choice of modulation. Thus, the maximal attainable key rate results from the optimization of Eq.~\eqref{eq:K_ind} over $\delta_B$ (or equivalently $\delta_E$) and is then seen to be a function of $\mathscr{E}$ only. From the security point of view, the worst-case scenario occurs when Eve's detection is shot-noise-limited, $\sigma_E^2 = \tau_E \bar{n}$, so that $\delta_E = \sqrt{\tau_E} (n_1 - n_0) / (2\sqrt{\bar{n}})$. On the other hand, for Bob the variance can be decomposed as $\sigma_B^2 = \tau_B \bar{n} + \sigma_{B,\textrm{ex}}^2$ with the first term corresponding to shot noise and the second to any additional source of noise. Note that due to the macroscopic pulse strength assumption, we may use the same variance value for both the $0$ and $1$ pulses that depends only on the average $\bar{n}$. Taking the above into account, the eavesdropper's advantage can be written as
\begin{equation}
\label{eq:eavesdropper_adv}
    \mathscr{E} = \frac{\tau_E}{\tau_B} \left(1  + \frac{\sigma_{B,\textrm{ex}}^2}{\tau_B \bar{n}}\right)
\end{equation}
and automatically accommodates increased values of Bob's noise. 

An approximate expression for the key rate is derived in~\cite{OKD} by expanding the integrand in Eq.~\eqref{eq:HBE} in the limit of $\mathscr{E} \gg 1$, indicating a strong advantage for Eve, which results in
\begin{equation}
\label{eq:KDDapprox}
    \mathsf{K}^{\text{DD}} (\mathscr{E} \gg 1) \approx \gamma \cdot \frac{\log_2{e}}{2 \mathscr{E}},
\end{equation}
where
\begin{equation}
\gamma =\max_{\delta \geq 0} \left \{ \delta^2 \left[ 2 - \int_{\mathbbm R} \mathrm{d}t\, \frac{e^{-(t-\delta)^2/2}}{\sqrt{2 \pi}}\, \frac{\sinh (2 \delta t) + 1}{\cosh^2 (\delta t)} \right] \right \}
\end{equation}
can be evaluated numerically to equal $\gamma \approx 0.4795$. Both the key rate values obtained by numerically optimizing the difference of Eqs.~\eqref{eq:HBA} and~\eqref{eq:HBE} over $\delta_B$, as well as the approximate expression~\eqref{eq:KDDapprox} for strong eavesdropping, are depicted in Fig.~\ref{fig2} as functions of the eavesdropper's advantage $\mathscr{E}$. 

\section{Coherent detection eavesdropping}
\label{sec:coh}

In the second quantization picture, the light pulses sent by Alice can be described by coherent states $\ket{\sqrt{n_a} e^{i \phi_a}}$, where $\sqrt{n_a}$ signifies the amplitude and $\phi_a$ the phase. Phase-insensitive direct detection performed by Bob on such states yields the photocount statistics discussed in the previous section, which in the macroscopic signal strength regime can be described by the variable $y_B$ defined in Eq.~\eqref{eq:yB_yE}. If, however, Eve possesses coherent detection capabilities, she may try to distinguish between the two states by measuring their optical phase-dependent quadratures instead of only the intensity, as a difference in phase increases the phase-space distance between the two states. From the point of view of Alice, it is therefore sensible to randomize the phase $\phi_a$ for every pulse. This requirement has been recognized as important for QKD security~\cite{Lo2007, CurrsLorenzo2023} and can be mitigated, for instance, with a phase modulator driven by quantum random number generation~\cite{Zhao2007} or gain-switched laser diodes~\cite{Lo2026}. Phase-randomized coherent states are characterized only by the photon number distribution identical to that of Sec.~\ref{sec:dd}, in which case coherent detection does not offer any advantage over direct detection and the security analysis therein remains applicable. On the other hand, if the phase of Alice's pulses is fixed, for example when they are carved with an electro-optic modulator from a narrowband continuous-wave laser output, it is important to ensure that in each individual transmission the choice of signal strength is uncorrelated with the phase, $\phi_0 = \phi_1$, so that the latter does not carry any information about the value of the variable $A$. In the pessimistic scenario in which Eve has the ability to sync to the pulses' phase, one may simply set $\phi_a = 0$ in the analysis and formulate her task as distinguishing between two coherent states $\ket{\sqrt{\tau_E n_0}}$ and $\ket{\sqrt{\tau_E n_1}}$ in each protocol round. Then, if Eve performs shot-noise-limited homodyne detection of the amplitude quadrature, she obtains a result $q_E$ distributed according to~\cite{Banaszek2020}
\begin{equation}\label{eq:homodyne_dist}
q_E|a \sim \mathcal{N} \left( \sqrt{2 \tau_E n_a}, 1/2 \right).
\end{equation}
Following Section~\ref{sec:dd}, one may define an equivalent rescaled variable
\begin{equation}
{y_E^{\text{coh}} = (q_E-\sqrt{2\tau_E}\bar{\alpha}) / \sqrt{1/2}}
\end{equation}
with $\bar{\alpha} = (\sqrt{n_0} + \sqrt{n_1})/2$. Note that the factor $(\sqrt{1/2})^{-1}$ follows from the the standard deviation of the conditional distribution. When conditioned on $A$, Eve's variable is seen to have a similar form to Eq.~\eqref{eq:yB_yE},
\begin{equation}
\label{eq:pE_coh}
    y_E^{\text{coh}}|a \sim \mathcal{N}\left((-1)^{1-a} \,\delta_E^{\text{coh}}, 1\right),
\end{equation}
but with a different modulation depth given by
\begin{equation}
\label{eq:delta_E_H}
    \delta_E^{\mathrm{coh}} = \sqrt{\tau_E} \left( \sqrt{n_1}-\sqrt{n_0} \right),
\end{equation}
calculated here as half of the difference between the peaks of $p(y_E^{\text{coh}}|n_0)$ and $p(y_E^{\text{coh}}|n_1)$. Therefore, the joint distribution $p(y_B, y_E^{\text{coh}})$ also assumes the same form as in Eq.~\eqref{eq:joint_DD_distribution} and leads to the same expression for the key rate, the only difference being that now Eve's modulation depth is $\delta_E^{\mathrm{coh}}$ instead of $\delta_E$. However, the two can be related by
\begin{align}
    \delta_E^{\mathrm{coh}} &= \delta_E \, \frac{\sqrt{2(n_0 + n_1)}}{\sqrt{n_0} + \sqrt{n_1}} \nonumber \\ &= \delta_E \left[1 + \frac{1}{32} \left(\frac{\Delta n}{\bar{n}}\right)^2 + \mathcal{O} \left(\left(\frac{\Delta n}{\bar{n}}\right)^4\right) \right]
\end{align}
with $\Delta n := n_1 - n_0$. Because we already assumed the input energy to be large $\bar{n} \gg 1$ and $n_0 \approx n_1$, to a good approximation we have $\delta_E^{\mathrm{coh}} \approx \delta_E$, which means that the key rate calculations for direct detection eavesdropping are reproduced exactly with coherent eavesdropping and one obtains to a good approximation $\mathsf{K}^{\text{coh}} = \mathsf{K}^{\text{DD}}$.
Thus, OKD retains the same level of security against coherent detection as against direct detection, provided that Alice does not encode any information in the phase quadrature.

The equivalence should in fact be expected since for states aligned across the $\phi_a = 0$ axis in phase space, direct detection effectively measures $q_E^2$. For large signal strengths the negative tail of the Gaussian distribution in Eq.~(\ref{eq:homodyne_dist}) is negligible and Eve observes only positive quadrature values which she can equivalently infer from direct detection results by taking a square root without any sign ambiguity, meaning both types of measurements provide the same information.

\section{Eavesdropping based on minimum-error state discrimination}
\label{sec:helstrom}
The natural next question is whether there are better ways for Eve to distinguish between the states $\ket{\sqrt{\tau_E n_a}}$. Indeed, in quantum detection and estimation theory the strategy characterized by the lowest probability of error in discriminating between two quantum states is given by the Helstrom measurement~\cite{Helstrom1969}. 

For the specific case of two coherent states $\ket{\alpha_a} = \ket{\sqrt{\tau_E n_a}}$, $a\in \{0,1\}$, this measurement can be performed with the Dolinar receiver~\cite{Dolinar1973} and yields the minimum discrimination error probability
\begin{align}
\label{eq:Perr}
P_{\text{err}}&=\frac{1}{2}\left( 1 - \sqrt{1 - |\!\braket{\alpha_0 | \alpha_1}\!|^2} \right) = \frac{1}{2}\left[ 1 - \sqrt{1 - e^{ - \left(\delta_E^\mathrm{coh}\right)^2 }} \right],
\end{align}
where $\left|\braket{\sqrt{\tau_E n_0} | \sqrt{\tau_E n_1}}\right|^2 = e^{-\tau_E (\sqrt{n_1}-\sqrt{n_0})^2 }$ and we used Eq.~\eqref{eq:delta_E_H}.

Let us now focus on the calculation of the key rate. When conditioned on $A=a$, the results of Bob, $y_B$, and Eve, $m_E$, are uncorrelated, and the corresponding joint probability distribution reads $p(a,y_B,m_E)=p(a)\,p(y_B|a)\,p(m_E|a)$ where $p(a)=1/2$ and $p(y_B|a)$ is still given by Eq.~\eqref{eq:yB_yE}. If Eve implements the Helstrom measurement with outcome $m_E \in \{0,1\}$, we can write
\begin{equation}
\label{eq:p_E_cond_a}
p(m_E|a) = \left\{\begin{array}{ll}
1-P_\text{err} & m_E=a\\[1ex]
P_\text{err} & m_E\ne a
\end{array}\right.
\end{equation}
and we can easily retrieve the conditional probability
\begin{equation}
p(y_B,m_E) = \frac12\sum_{a} p(y_B|a)\,p(m_E|a).
\end{equation}
Eve's outcome being discrete, the expression for the conditional entropy now reads
\begin{equation}
\label{eq:HBE_Helstrom}
    H(B|E) = - \frac12\sum_{m_E}\int_{\mathbbm R} \mathrm{d} y_B \, p(y_B|m_E) \log_2 p(y_B|m_E),
\end{equation}
where we used $p(y_B, m_E) =p(y_B|m_E) p(m_E)$ and $p(m_E) = 1/2$. The explicit analytical expression of Eq.~(\ref{eq:HBE_Helstrom}) is clumsy and
results in a lengthy integral which is not analytically solvable. However, to obtain a closed-form approximate result, one may again enforce the limit of large eavesdropper's advantage, $\mathscr{E} \gg 1$ or, equivalently, $\delta_B \ll 1$ due to Eq.~\eqref{eq:deltas_relation}. If one expands the probabilities up to second order in $\delta_B$, Eq.~\eqref{eq:HBE_Helstrom} reduces to
\begin{align}
    &H(B|E) = \frac{1}{2\ln 2} \int_{\mathbbm R} \mathrm{d}y_B \;
    \frac{e^{-y_B^2 / 2}}{\sqrt{2\pi}} \Bigg\{  \left(y_B^2+\ln 2\pi
   \right) \nonumber \\ 
   &  + \delta_B^2 \, \frac{
   y_B^4 + y_B^2 \left[\ln 2\pi -5 +2 \,e^{-({\delta_E^\mathrm{coh}})^2}\right] + 2 - \ln 2\pi}{2 } \Bigg\}
\end{align}
and can now be evaluated analytically. By combining the result with Eq.~\eqref{eq:HBA} and plugging $\delta_B = \delta_E^\mathrm{coh} / \sqrt{\mathscr{E}}$, the key rate in Eq.~\eqref{eq:K_ind} for the case of Eve performing optimal Helstrom measurements greatly simplifies to
\begin{equation}
\label{eq:K_Helstrom_approx}
   \mathsf{K}^\text{Helstrom}(\delta_E^\mathrm{coh}, \mathscr{E} \gg 1)= \frac{\log_2{e}}{2\mathscr{E}}\cdot {\delta_E^\mathrm{coh}}^2 e^{-({\delta_E^\mathrm{coh}})^2}.
\end{equation}
The expression in Eq.~\eqref{eq:K_Helstrom_approx} can now be maximized over $\delta_E^\mathrm{coh}$, which is equivalent to Alice choosing optimal values of energy $n_0$ and $n_1$ for the light pulses. This gives us a rather elegant result for the key rate secure against passive eavesdropping that depends only on $\mathscr{E}$:
\begin{equation}
\label{eq:K_Helstrom_approx_opt}
\mathsf{K}^\text{Helstrom}(\mathscr{E} \gg 1) \approx \frac{1}{e} \cdot \frac{\log_2{e}}{2 \mathscr{E}}.
\end{equation}
Note that the key rate in Eq.~\eqref{eq:K_Helstrom_approx_opt} exhibits the same scaling with $\mathscr{E}$ as for direct/coherent detection in Eq.~\eqref{eq:KDDapprox} but with a lower multiplicative factor due to the Helstrom measurement allowing for better state discrimination by Eve. Both the full numerical solution and approximate analytical key rate values are plotted in Fig.~\ref{fig2}.

\section{Collective eavesdropping}
\label{sec:holevo}

\begin{figure*}[t]
\includegraphics[width=0.95\textwidth]{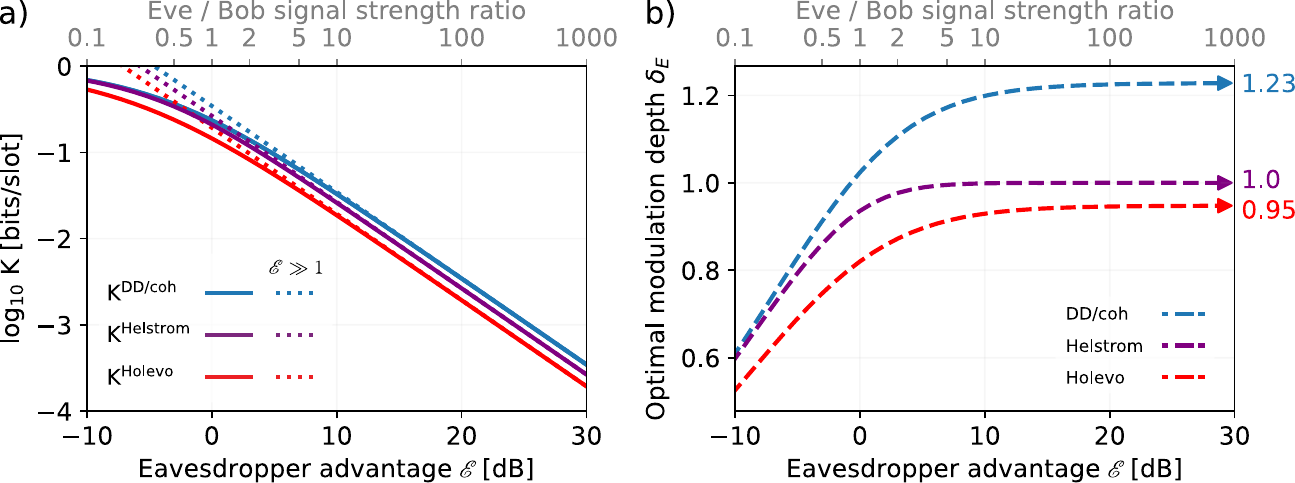}
\caption{a) Key generation rates as a function of the eavesdropper's advantage $\mathscr{E}$ for the case of passive eavesdropping with: direct/coherent detection (blue, Sec.~\ref{sec:dd}~\&~\ref{sec:coh}); optimal state discrimination via Helstrom measurements (purple, Sec.~\ref{sec:helstrom}); optimal measurements saturating the Holevo bound (red, Sec.~\ref{sec:holevo}). Dotted lines indicate approximations derived in text for the strong eavesdropping limit, $\mathscr{E} \gg 1$. The upper axis indicates the ratio of Eve's and Bob's signal strength if Bob's detection is shot-noise limited, in which case $\mathscr{E} = \tau_E / \tau_B$ according to Eq.~\eqref{eq:eavesdropper_adv}. b) Values of the modulation depth $\delta_E$ that maximize the key rate for the three considered passive eavesdropping scenarios as a function of the eavesdropper's advantage. Arrows indicate the optimal values that maximize the corresponding asymptotic key rate expressions in Eq.~\eqref{eq:KDDapprox},
\eqref{eq:K_Helstrom_approx_opt}, and \eqref{eq:KHolevoapprox}. }
\label{fig2}
\end{figure*}

From the point of view of Eve, minimizing the key rate $\mathsf{K}$ in Eq.~\eqref{eq:K_ind} entails maximizing her correlation with Bob's outcomes, quantified by the mutual information $I(B\,;E)$. Quantum theory provides a theoretical upper bound to this value given by the Holevo quantity $\chi(B\,;E)$~\cite{Holevo1973,Banaszek2020}, resulting in a generalized key rate expression
\begin{equation}
\label{eq:KHol}
    \mathsf{K}^\text{Holevo} = \max \{I(A\,;B) - \chi(B\,;E), \; 0\}.
\end{equation}
On the one hand, here the mutual information $I(A\,;B)$ between Alice and Bob can be expressed as~\cite{OKD}
\begin{equation}
    I(A\,;B) = \delta_B^2 \log_2 e - \int_{\mathbbm R} \mathrm{d}t \frac{e^{-(t - \delta_B)^2 / 2}}{\sqrt{2 \pi}}\, \log_2 \left[ \cosh (\delta_B t) \right],
\end{equation}
where $\delta_B$ has been introduced in the previous sections.
On the other hand, the Holevo quantity $\chi(B\,;E)$ between Bob and Eve quantifies the maximal amount of information that Eve can learn about Bob's classical variable $B$ from the received quantum states, optimized over all possible quantum operations that she can perform. This includes collective attacks in which she may store the collected states in a quantum memory and measure them jointly later, which can in theory give her more information than individual round-by-round measurements~\cite{Preskill2004}. The Holevo quantity is defined as
\begin{equation}
\label{eq:chiBE}
    \chi(B\,;E) = S\left[ \rho_E\right] - \int_{\mathbb R} \mathrm{d}y_B \, p(y_B) S[\rho_{E|B=y_B}],
\end{equation}
where $S[\rho] = -\text{Tr} \left( \rho \log_2 \rho \right)$ is the von Neumann entropy of a quantum state $\rho$,
\begin{equation}
\label{rho:E}
\rho_E = \frac12 \sum_a\ketbra{\sqrt{\tau_E n_a}}{\sqrt{\tau_E n_a}}_E,
\end{equation}
is the overall state received by Eve, which is a mixture of two coherent states, whereas $\rho_{E|B=y_B}$ is her state conditioned on Bob's outcome $y_B$.

In order to calculate $\chi(B\,;E)$ we start from the tripartite state after Alice's transmission, that, since we are in the presence of coherent states, can be written as \cite{olivares2021}
\begin{equation}\label{rho:ABE}
    \rho_{ABE}^{\rm (in)} = \frac12 \sum_a
    \rho^{(A)}_a
    \otimes\rho^{(B)}_{a}
    \otimes\rho^{(E)}_{a},
\end{equation}
where $\rho^{(K)}_{a}= \ketbra{\sqrt{\tau_K n_a}}{\sqrt{\tau_K n_a}}_K$ is the state received by $K\in\{B,E\}$ given the input signal $\rho^{(A)}_a = \ketbra{a}{a}_A$, with $a\in\{0,1\}$ as usual.
After Bob performs direct detection, the state transforms to
\begin{equation}
    \rho_{ABE} = \frac12 \sum_{a}  \rho^{(A)}_a \otimes \rho_{B|A=a} \otimes\rho^{(E)}_{a},
\end{equation}
where his part is now a classical mixture of outcomes $y_B$ distributed according to Eq.~\eqref{eq:bob_eve_variables} and denoted by orthogonal states $\ket{y_B}$,
\begin{equation}
\label{eq:Bob_cond_state}
    \rho_{B|A=a} = \int_{\mathbbm R}  \mathrm{d}y_B \, \frac{e^{- [y_B + (-1)^a \delta_B]^2/2}}{\sqrt{2 \pi}}\, \ketbra{y_B}{y_B}.
\end{equation}
Thereafter, we have the following joint state of Bob and Eve
\begin{equation}
\rho_{BE} =\text{Tr}_A (\rho_{ABE})=\int_{\mathbbm R} \mathrm{d}y_B \, p(y_B) \ketbra{y_B}{y_B} \otimes \rho_{E|B=y_B},
\end{equation}
where
\begin{equation}
p(y_B) = \frac12\sum_a \frac{\displaystyle e^{- [y_B + (-1)^a \delta_B]^2/2}}{\sqrt{2 \pi}}
\end{equation}
and
\begin{align}
\rho_{E|B=y_B} &=
\frac{\displaystyle\sum_{a} e^{- [y_B + (-1)^a \delta_B]^2/2}\, \rho^{(E)}_a
}{\displaystyle\sum_a e^{- [y_B + (-1)^a \delta_B]^2/2}}.
\label{eq:rho_E_cond_B}
\end{align}

In summary, both $\rho_E$ and $\rho_{E|B=y_B}$ appearing in Eq.~(\ref{eq:chiBE}) are mixtures of two coherent states. Since the von Neumann entropy of a general mixture of two coherent states $\rho = p \ketbra{\alpha}{\alpha} + (1-p) \ketbra{\beta}{\beta}$ is given by 
\begin{equation}
\label{eq:vonNeumannSol}
S[\rho] = h\left[ \frac{1-\sqrt{1 - 4p(1-p)(1-|\!\braket{\alpha|\beta}\!|^2)}}{2}  \right ]\!,
\end{equation}
where $h[x]=-x\log_2x-(1-x)\log_2(1-x)$ is the binary entropy function, one can straightforwardly calculate the entropies $S\left[ \rho_E \right]$ and $S[\rho_{E|B=y_B}]$ in Eq.~\eqref{eq:chiBE}.

Given $I(A\,;B)$ and $\chi(B\,;E)$, the key rate expression can now be optimized numerically; nevertheless, one may again determine an approximate expression valid for strong eavesdropping by expanding the integrands up to second order in $\delta_B$ and optimizing over $\delta_E$. The result, depicted in Fig.~\ref{fig2}, is seen to follow the same scaling with $\mathscr{E}$ as in the previous cases of Sec.~\ref{sec:dd}, \ref{sec:coh}, and \ref{sec:helstrom},
\begin{align}
    \mathsf{K}^{\text{Holevo}}(\mathscr{E} \gg 1) \approx \chi \cdot \frac{\log_2{ e}}{2 \mathscr{E}},
\end{align}
with
\begin{align}
    \chi &= \max_{\delta \geq 0} \left \{ \delta^2 \left[ 1 - 2 \coth^{-1} \! \left( e^{\delta^2 / 2} \right) \sinh \left( \delta^2 / 2 \right) \right] \right \} \approx 0.2683.
\label{eq:KHolevoapprox}
\end{align}
Crucially, the key rate is further reduced by a multiplicative factor when compared to direct/coherent detection and the Helstrom measurement, but the scaling with $\mathscr{E}$ still remains the same.

\section{Discussion}
\label{sec:conclusion}
We have analyzed the asymptotic information-theoretic key generation rates attainable by optical key distribution, assuming the regime of macroscopic signal strengths and Gaussian photocount statistics, in the presence of a malicious passive eavesdropper with general capabilities. These include coherent eavesdropping with homodyne detection, shown to be equivalent to the direct detection scenario, and generalized quantum strategies: the Helstrom measurement minimizing the state discrimination error probability and measurements saturating the ultimate Holevo bound on accessible information. In both of the quantum-enhanced cases the attainable key rate values are reduced, but only by multiplicative factors compared to direct and coherent detection, while following the same asymptotic scaling with a parameter quantifying the eavesdropper's advantage. In conclusion, granting the passive eavesdropper enhanced quantum capabilities does not prohibit secure key distribution between the honest parties using the Optical Key Distribution protocol. Remarkably, the crucial restriction of passive eavesdropping imposed on Eve in OKD is enough to provide strong security guarantees, even if she is allowed to perform arbitrary quantum operations on her intercepted signal.

It is interesting to note that in the regime of strong eavesdropping, $\mathscr{E} \gg 1$, we found the same scaling $1/{\mathscr E}$ for all the considered strategies of Eve, with more powerful quantum eavesdropping resulting only in lower multiplicative prefactors in the key rate formulas. Intuitively, this may result from the following interplay. First of all, the passive eavesdropping model and the use of coherent states as information carriers result in limited correlation between Bob's and Eve's outcomes given the value of $a$, as although Eve can improve her state discrimination power, she cannot influence the random result obtained by Bob. Second of all, we observe in all the cases that it is optimal for Alice to choose a modulation depth on the order of $\delta_E \approx 1$, equivalent to the phase space separation of Eve's received coherent states. In this regime, quantum-enhanced measurements are not expected to qualitatively change Eve's distinguishability, in contrast to the limit of vanishing state separation where they offer a fundamental advantage over classical detection~\cite{Banaszek2020}. To further elucidate this phenomenon, more research into the interplay of quantum and classical protocols is needed.

There are multiple other interesting directions to explore in relation to this work. While this is a fundamental study of the protocol limits under quantum passive eavesdropping, it would be interesting to see how practical considerations such as finite-size effects, limited reconciliation efficiency, or the use of hard decoding~\cite{banaszek2021optimization} would influence the attainable key rates. In terms of phase control discussed in the section on coherent eavesdropping, one could study the effect of imperfect phase randomization of Alice's states or information leakage through a residual correlation between phase and signal strength. One could also consider the use of quantum signals by Alice, such as displaced squeezed states, or other photon-number-resolving-based strategies for Eve like the weak-field homodyne detection~\cite{notarnicola:25}. Finally, equipping Bob with quantum-enhanced capabilities could improve the attainable rates, but would move the protocol away from its attractive simplicity and integrability with modern communications infrastructure.

\begin{acknowledgments}
This work was supported by the Polish Ministry of Education and Science under the “Quantum strategies in communication through noisy optical channels” project no. PN/01/0204/2022 carried out within the “Pearls of Science” program. This work was supported by the Foundation for Polish Science under the ”Quantum Optical Technologies” project carried out within the International Research Agendas programme co-financed by the European Union under the European Regional Development Fund.
\end{acknowledgments}

\section*{Appendix. Validation of approximations and small signal strength behaviour}
\label{sec:appendix}
In the main text we derived compact key rate expressions owing to the assumption of macroscopic pulse intensities measured by Bob's and Eve's detectors, $\tau_{B/E} \, n_a \gg 1$, which allowed us to approximate the Poissonian statistics of photodetection with normal distributions. Moreover, we constricted Alice's modulation depth to be small, meaning that her two transmitted signal strengths are similar, $n_0 \approx n_1$, in order to prevent unambiguous detection of the bit value by Bob and Eve. This and the fact that photocount variance generally scales proportionally with the signal strength prompted another simplifying approximation in which we set the same variance value for both the $0$ and $1$ pulse as a function of $\bar{n} = (n_0 + n_1)/2$ only. To assess the validity of these simplifications, in this Appendix we compare the main text results with an exact numerical treatment. By doing so, we also show how the key rates behave in the regime of small signal strengths. Below we first explain the necessary changes to the main text derivations if one were to drop the above approximations. Next we plot the comparison of approximated and exact calculations and show that the two converge quickly with increasing signal strength. 

The protocol structure remains the same: in each OKD round Alice transmits one of two coherent states with an average number of photons $n_a \in \{n_0, n_1\}$ and equal probability $p(a) = 1/2$, after which Bob performs direct detection of the received pulses attenuated by transmission $\tau_B$. At the shot noise level his number of photocounts $k_B$ is then distributed according to a Poisson distribution with mean $\tau_B n_a$ dependent on the value of $a$. One may consider Bob's detection to be noisy and have his outcome be a sum of signal and noise random variables. A natural choice is to model the latter also as an independent Poisson process with mean $\sigma^2_{B,\mathrm{ex}}$, which is characteristic of direct detection in the presence of dark counts or thermal radiation. As a result, Bob's exact outcome is distributed as $k_B^{\mathrm{P}} | a \sim \mathrm{Pois}(\tau_B n_a + \sigma^2_{B,\mathrm{ex}})$ and its probability mass function reads
\begin{equation}
    p(k_B^{\mathrm{P}}|a) = \frac{\exp(-\tau_B n_a - \sigma^2_{B,\mathrm{ex}}) \cdot [\tau_B n_a + \sigma^2_{B,\mathrm{ex}}]^{k_B^{\mathrm{P}}}}{k_B^{\mathrm{P}}!}
\end{equation}
with both the mean and variance equal to $\tau_B n_a + \sigma^2_{B,\mathrm{ex}}$. Depending on the chosen scenario, other forms of contributing noise could also be considered here, which would necessitate a different form of Bob's outcome distribution. Note that in the main text, excess noise was modeled simply by adding a factor $\sigma^2_{B,\mathrm{ex}}$ to the Gaussian outcome variance.

Apart from Alice and Bob, the protocol assumes the presence of a passive eavesdropper Eve measuring pulses attenuated by transmission $\tau_E$, who we have equipped in the main text with increasingly enhanced detection capabilities. As is customary in security analysis, Eve's detection has no excess noise to cover for the worst-case scenario. If she performs direct detection, her exact outcome is now $k_E^{\mathrm{P}} \sim \mathrm{Pois}(\tau_E n_a)$, i.e., a shot-noise limited Poisson random variable with mean and variance dependent on $a$. If Eve performs coherent detection as in Sec.~\ref{sec:coh} or minimum-error state discrimination as in Sec.~\ref{sec:helstrom}, her outcome distributions in Eq.~\eqref{eq:homodyne_dist} and Eq.~\eqref{eq:p_E_cond_a} do not change, as we did not need to enforce any approximation for these cases. Finally, if Eve is equipped with collective eavesdropping as in Sec.~\ref{sec:holevo}, in the derivation one needs to replace Bob's conditional state in Eq.~\eqref{eq:Bob_cond_state} with
\begin{equation}
    \rho_{B|A=a} = \sum_{k_B^{\mathrm{P}} = 0}^{\infty} p(k_B^{\mathrm{P}}|a) \ketbra{k_B^{\mathrm{P}}}{k_B^{\mathrm{P}}}
\end{equation}
and update the following expressions of Sec.~\ref{sec:holevo} accordingly.

\begin{figure*}[t]
\centering
\includegraphics[width=\textwidth]{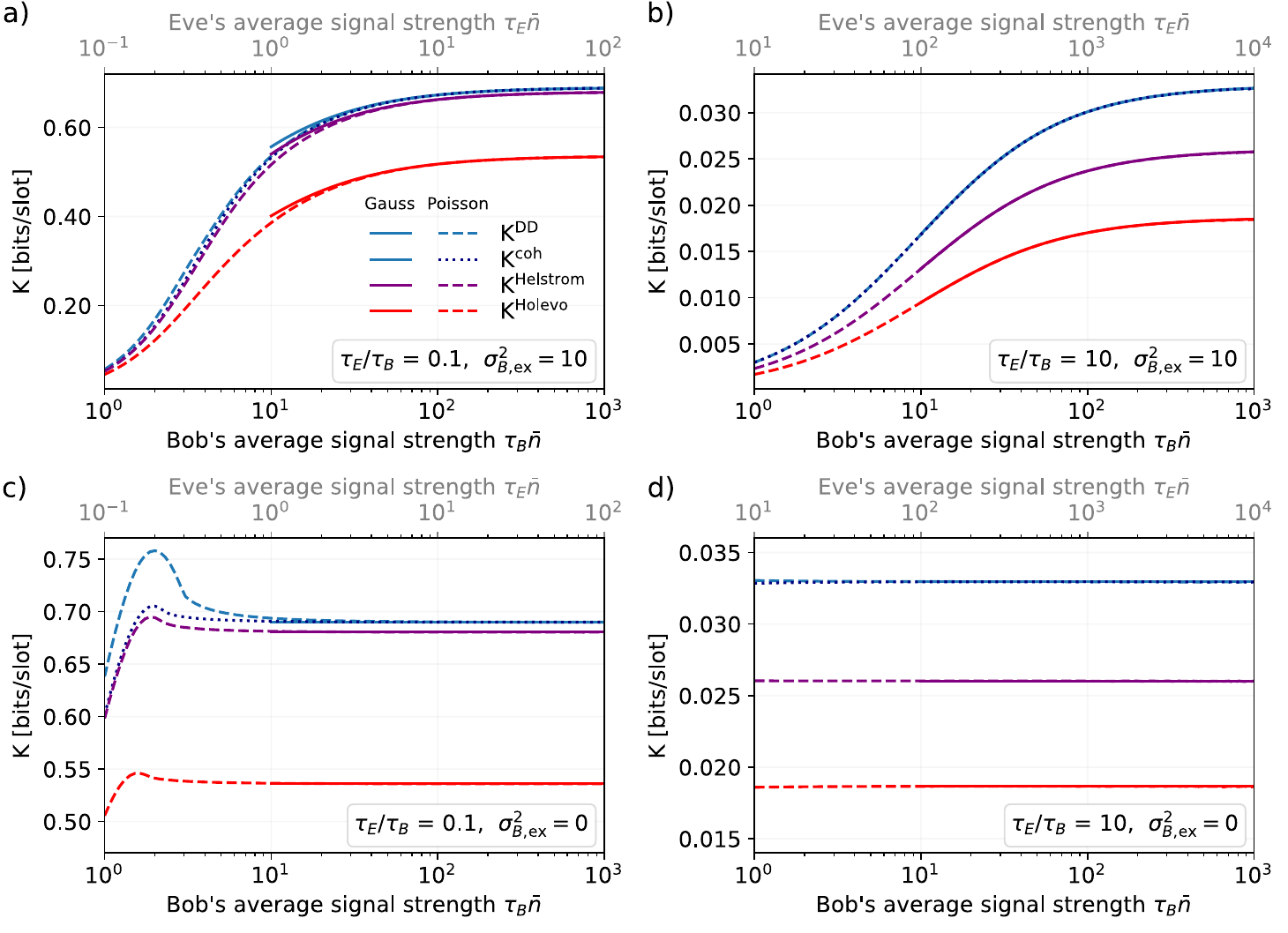}
\caption{Comparison of key generation rates of the model derived in the main text under macroscopic pulse strength and small modulation depth approximations (solid lines) with an exact model delineated in the Appendix (dashed and dotted lines) for select representative cases of the $\tau_E / \tau_B$ ratio and excess noise $\sigma^2_{B,\mathrm{ex}}$.}
\label{fig_appendix}
\end{figure*}

Without the opportunity for simplification of the formulas through approximations, one has to express the key generation rates as functions of the exact outcome probability distributions and calculate them numerically. For the sake of completeness, let us list the full expressions to be used. For individual attacks in Eq.~\eqref{eq:K_ind} one now takes
\begin{equation}
    H(B|A) = - \sum_{a} p(a)\! \left(\!\sum_{k_B^{\mathrm{P}} = 0}^{\infty} p(k_B^{\mathrm{P}}|a) \log_2 p(k_B^{\mathrm{P}}|a)\! \right)
\end{equation}
and, if Eve performs direct detection,
\begin{equation}
    H(B|E) = -\!\!\!\sum_{k_E^{\mathrm{P}}, k_B^{\mathrm{P}}=0}^{\infty}\hspace{-0.6em}p(k_B^{\mathrm{P}}, k_E^{\mathrm{P}}) \log_2 \frac{p(k_B^{\mathrm{P}}, k_E^{\mathrm{P}})}{p(k_E^{\mathrm{P}})}\IEEEeqnarraynumspace
\end{equation}
with $p(k_B^{\mathrm{P}}, k_E^{\mathrm{P}}) = \sum_{a} p(a) p(k_B^{\mathrm{P}}|a) p(k_E^{\mathrm{P}}|a)$ and $p(k_E^{\mathrm{P}}) = \sum_{a} p(a) p(k_E^{\mathrm{P}}|a)$. If Eve performs the Helstrom measurement, her outcome is still discrete but now has two values $m_E \in \{0,1\}$ distributed as in Eq.~\eqref{eq:p_E_cond_a} and one needs to substitute $k_E^{\mathrm{P}}$ for $m_E$ in the above formula. For Eve with coherent detection capabilities, her outcome is continuous and her conditional probability $p(y_E^{\mathrm{coh}}|a)$ is given by Eq.~\eqref{eq:pE_coh}, in which case
\begin{equation}
    H(B|E) = - \! \sum_{k_B^{\mathrm{P}}=0}^{\infty} \int_{\mathbbm R} \mathrm{d}y_E^{\mathrm{coh}} p(k_B^{\mathrm{P}}, y_E^{\mathrm{coh}}) \log_2 \frac{p(k_B^{\mathrm{P}}, y_E^{\mathrm{coh}})}{p(y_E^{\mathrm{coh}})}
\end{equation}
with $p(k_B^{\mathrm{P}}, y_E^{\mathrm{coh}}) = \sum_{a} p(a) p(k_B^{\mathrm{P}}|a) p(y_E^{\mathrm{coh}}|a)$ and $p(y_E^{\mathrm{coh}}) = \sum_{a} p(a) p(y_E^{\mathrm{coh}}|a)$. Finally, for collective attacks in Eq.~\eqref{eq:KHol} one resorts to the full expression
\begin{equation}
    I(A\,;B) = \sum_{a}\sum_{k_B^{\mathrm{P}}=0}^{\infty} p(a, k_B^{\mathrm{P}}) \log_2 \left( \frac{p(a, k_B^{\mathrm{P}})}{p(a) p(k_B^{\mathrm{P}})} \right)
\end{equation}
with $p(a, k_B^{\mathrm{P}}) = p(a) p(k_B^{\mathrm{P}}|a)$ and $p(k_B) = \sum_{a} p(a,k_B^{\mathrm{P}})$, while the calculation of $\chi(B\,;E)$ proceeds again as a difference of two von Neuman entropies of coherent state mixtures, only with Bob's exact outcome probability distribution $p(k_B^{\mathrm{P}}|a)$ used in place of the approximated Gaussian one.

The lack of approximations makes life harder in two ways. First of all, exact calculations of the above expressions call for infinite sums and integrals to be computed. To circumvent that, we proceed each time by calculating these numerically with a high enough upper limit and precision that raising them further does not influence the results anymore. Second of all, while previously key rate expressions were derived as functions of a single eavesdropper's advantage parameter $\mathscr{E}$, now they are functions of individual parameters including transmissions $\tau_B$, $\tau_E$, excess noise $\sigma^2_{B,\mathrm{ex}}$, and transmitted pulse intensities $n_0$ and $n_1$. 

Fig.~\ref{fig_appendix} plots the attainable key generation rates $\mathsf{K}$ secure against passive eavesdropping as a function of Bob's average signal strength $\tau_B \bar{n}$ for four representative cases of either Eve highly attenuated with respect to Bob, $\tau_E / \tau_B = 0.1$, or Bob with respect to Eve, $\tau_E / \tau_B = 10$, and either nonzero excess noise in Bob's measurement, $\sigma^2_{B,\mathrm{ex}} = 10$, or noiseless direct detection, $\sigma^2_{B,\mathrm{ex}} = 0$. The exact values of key generation rates calculated according to the prescription in this Appendix are plotted with dashed and dotted lines, where for each $\tau_B \bar{n}$, the values of $\tau_B n_0$ and $\tau_B n_1$ are optimized for highest $\mathsf{K}$. Additionally, the approximate key rate formulas derived in the main text under the macroscopic pulse strength and small modulation depth assumptions are plotted with solid lines from $\tau_B \bar{n} = 10$ onwards, as from roughly this value one can generally start to approximate Poissonian distributions with Gaussians. In this case, for each $\tau_B \bar{n}$ and the values of $\tau_E / \tau_B$ and $\sigma^2_{B,\mathrm{ex}}$ we calculate $\mathscr{E}$ according to Eq.~\eqref{eq:eavesdropper_adv} and plot the key rates derived as functions of $\mathscr{E}$ in Sec.~\ref{sec:dd} through~\ref{sec:holevo}. In subfigures a) and b) the nonzero excess-noise-related penalty to $\mathscr{E}$ in Eq.~\eqref{eq:eavesdropper_adv} ranges from high to small throughout the plotted range. In the noiseless case of subfigures c) and d), $\mathscr{E}$ no longer depends on $\tau_B \bar{n}$, which results in constant values for the solid Gaussian lines.

The exact and approximate values are seen to closely coincide with each other already very well at medium photon numbers, $\tau_{B/E}\, \bar{n} \approx 10$, and reach excellent agreement with increasing $\tau_B \bar{n}$. This validates the main text assumptions and shows that the key rate formulas derived therein are quite general and fully applicable. In addition to showing that, the plot elucidates the small signal strength behaviour of OKD with enhanced passive eavesdropping. Crucially, if a constant value of excess noise is present, it is seen that key rates generally increase with signal strength due to the corresponding decrease of the eavesdropper's advantage $\mathscr{E}$, which is at its lowest when shot noise dominates over excess noise in Bob's measurement.

\bibliography{bibliography}

\end{document}